\documentclass[twocolumn]{aastex7}

\usepackage[utf8]{inputenc}
\usepackage{amsfonts}
\usepackage{amsmath}
\usepackage{amssymb}
\usepackage{amsthm}
\usepackage{graphicx}
\usepackage{tikz}
\usepackage{xcolor}
\usepackage{natbib}
\usepackage{colortbl}
\usepackage{booktabs}
\usepackage{multirow}
\usepackage{color,soul}
\usepackage{ulem}
\usepackage{appendix}
\usepackage{comment}
\usepackage{lmodern}
\usepackage{orcidlink}
\usepackage{fix-cm}
\usepackage{hyperref}
\usepackage{etoolbox}

\newbool{editsshow}
\setbool{editsshow}{false} 

\newcommand{\REVNEW}[1]{\ifbool{editsshow}{\textbf{#1}}{#1}}
\newcommand{\REVDEL}[1]{\ifbool{editsshow}{\st{#1}}{}}

\usetikzlibrary{shapes.geometric, arrows, positioning}

\def \nsv {n \Sigma v}
\def \MBH {M_\bullet}
\def \mmm {\left( \frac{\MBH}{10^6 M_\odot} \right)}
\def \inpc#1 {\left( \frac{#1}{1\text{ pc}} \right)}
\newcommand{\parfrac}[2]{\left( \frac{#1}{#2} \right)}

\accepted{June 24, 2025}
\submitjournal{ApJ: \href{https://doi.org/10.3847/1538-4357/ade878}{https://doi.org/10.3847/1538-4357/ade878}}

\begin{document}

\title{The unreasonable effectiveness of the $n \Sigma v$ approximation}

\author[0000-0003-4105-3443]{Elisha Modelevsky}
\affiliation{Racah Institute of Physics, The Hebrew University, 91904, Jerusalem, Israel}
\email{elisha.modelevsky@mail.huji.ac.il}

\author[0000-0002-4337-9458]{Nicholas C. Stone}
\affiliation{Department of Astronomy, University of Wisconsin, Madison, WI, 53706}
\affiliation{Racah Institute of Physics, The Hebrew University, 91904, Jerusalem, Israel}
\email{nicholas.stone@mail.huji.ac.il}

\author[0000-0002-1084-3656]{Re'em Sari}
\affiliation{Racah Institute of Physics, The Hebrew University, 91904, Jerusalem, Israel}
\email{sari@phys.huji.ac.il}

\correspondingauthor{Elisha Modelevsky}
\email{elisha.modelevsky@mail.huji.ac.il}

\begin{abstract}

    In kinetic theory, the classic $n\Sigma v$ approach calculates the rate of particle interactions from local quantities: the number density of particles $n$, the cross-section $\Sigma$, and the average relative speed $v$.
    In stellar dynamics, this formula is often applied to problems in collisional (i.e. dense) environments such as globular and nuclear star clusters, where blue stragglers, tidal capture binaries, binary ionizations, and micro-tidal disruptions arise from rare close encounters.
    The local $n\Sigma v$ approach implicitly assumes the ergodic hypothesis, which is not well motivated for the densest star systems in the Universe.
    In the centers of globular and nuclear star clusters, orbits close into 1D ellipses because of the degeneracy of the potential (either Keplerian or harmonic).
    We find that the interaction rate in perfectly Keplerian or harmonic potentials is determined by a global quantity -- the number of orbital intersections -- and that this rate can be far lower or higher than the ergodic $n\Sigma v$ estimate.
    However, we find that in most astrophysical systems, deviations from a perfectly Keplerian or harmonic potential (due to e.g. granularity or extended mass) trigger sufficient orbital precession to recover the $n\Sigma v$ interaction rate.
    Astrophysically relevant failures of the $n\Sigma v$ approach only seem to occur for tightly bound stars orbiting intermediate-mass black holes, or for the high-mass end of collisional cascades in certain debris disks.
    
\end{abstract}


\section{Introduction}
\label{sec:intro}

Dense star systems are often termed ``collisional'' because of the dominant role that pairwise gravitational scatterings play in determining their bulk evolution.
However, these star clusters are also collisional in a second sense: their high density leads to a significant rate of close encounters between stellar-mass objects, including (but not limited to) physical collisions.
These close encounters can shape the properties of cluster members: in globular clusters, non-destructive star-star collisions produce easily identifiable blue stragglers \citep{HillsDay76, Bailyn95}, while in the dynamically ``hotter'' environments of galactic nuclei, physical collisions between pairs of stars may produce luminous transients \citep{Balberg+13, BalbergYassur23, Brutman+24} or destroy red giant envelopes \citep{Dale+09}.
Likewise, close star-binary interactions ionize wide binaries, harden tight binaries, and preferentially swap light stars out of pre-existing binary systems \citep{Heggie75, GoodmanHut89, Ivanova+05}.

Close encounters between combinations of stars and stellar mass compact objects may also lead to the production of interesting high-energy astrophysical sources.
Observations have long established that cataclysmic variables and neutron star X-ray binaries are massively over-represented in the dense cores of globular clusters \citep{Clark75, Grindlay+01, Pooley+03}, while a similar over-representation has more recently been identified for black hole X-ray binaries in the Milky Way's nuclear star cluster \citep{Hailey+18}.
The origins of these overabundances are likely dynamical and linked to frequent close encounters, although past models have debated the relative importance of two-body tidal captures \citep{Fabian+75, PressTeukolsky77, Generozov+18} and three-body binary-single scatterings \citep{Hills76, Ivanova+06, Ivanova+08, Kremer+20}.
The discoveries of gravitational wave emission from binary black hole mergers have also drawn much attention to dynamical formation of binary black holes via repeated binary-single scatterings in globular \citep{PortegiesZwartMcMillan00, Rodriguez+16} and nuclear \citep{AntoniniRasio16} star clusters.

All of these interesting astrophysical phenomena are driven by close encounters of particles in dense environments, and their rates have all been calculated in past works using the standard ``$n\Sigma v$'' kinetic approach.
Specifically, in a gas with particle number density $n$, the rate of collisions per particle is $\nsv$, where $\Sigma$ is the collision cross-section and 
$v$ is the mean relative speed of particles in the gas.
This statistical approach is widely used in gas and plasma kinetic theory, but also appears naturally well-suited for estimating encounter rates in stellar kinetic theory, and appears to be in good agreement with direct N-body simulations in those contexts for which it has been tested \citep{Reinoso+22}.

However, a crucial assumption behind the $\nsv$ formalism is \REVDEL{the uniformity of particles in space} \REVNEW{that the probability of encountering a particle at any given time is proportional to the density distribution $n$}.
In essence, this is the ergodic hypothesis -- all microstates are equiprobable over long periods of time; or in another formulation, long time averages are equivalent to averages over the statistical ensemble.
While the ergodic hypothesis is suitable for a chaotic gas (Boltzmann's ``Stosszahlansatz''), it will not be correct for every multi-particle system.
In systems where particles move in a degenerate potential, particle trajectories may be confined to restricted regions in phase space, and some regions in space will never be occupied (for a given realization of the system).

In this article, we will focus on three types of astrophysical potentials, with increasing levels of degeneracy:
\begin{enumerate}
    \item \underline{Spherically symmetric potentials}: trajectories are confined to a single plane of motion.
    \item \underline{Kepler potential}: trajectories are closed planar orbits.
    \item \underline{Harmonic potential}: trajectories are closed planar orbits, and {\it all} trajectories have the same period.
\end{enumerate}
The Kepler potential ($\Phi \propto r^{-1}$) and the harmonic potential ($\Phi \propto r^2$) are the only two spherically symmetric potentials that exhibit closed orbits \citep{bertrand1873,chin_2015}, so this is a complete set of degenerate potentials.

Although they may appear fine-tuned, these potentials are accurate (sometimes highly accurate) approximations for many multi-particle systems in astrophysics. 
Planetary orbits in the solar system are governed by the Sun's Kepler potential, just as stellar trajectories in the vicinity of a supermassive black hole are governed by its Kepler potential.
Near the centers of globular clusters and nuclear star clusters (at least, those lacking a central massive black hole), stars will relax into a constant-density core, creating a harmonic gravitational potential \citep{SpitzerHart71}.
Further away from the center of such clusters, the gravitational potential is not harmonic, but is still spherically symmetric.
Rather than representing some sort of unusual edge case, {\it these degenerate potentials describe the densest star systems in the Universe}, highlighting their importance  for a full understanding of astrophysical close encounter rates.

\REVNEW{
A related effect in degenerate potentials is \emph{resonant relaxation} \citep{rauch1996}, an increase in the relaxation rate due to stars remaining in the same orbit and exerting a persistent torque on other stars' orbits, as opposed to random impulses from uncorrelated encounters.
There are two reasons our analysis of collision rates differs from resonant relaxation:
first, collisions have a finite cross-section, so for a given set of orbits, only some (if any) of the stars' orbits interact;
second, collisions are usually destructive, which prevents persistent interactions.
}

In section \ref{sec:dynamics}, we study how the collision rate in such degenerate systems differs from ergodic systems.
In section \ref{sec:role_of_prec} we examine how precession (which removes the degeneracy) affects the collision rate.
Finally, in section \ref{sec:examples}, we apply these results to realistic astrophysical systems.

\section{Collision rate in degenerate systems}
\label{sec:dynamics}

\REVNEW{
In this section we examine how the collision rates behave in spherical, Kepler and harmonic potentials.
Let us clearly define the problem we solve.
Suppose we have a stellar phase-space distribution $f(\vec{r},\vec{v})$ with a corresponding density $n(\vec{r})=\int{\rm d}\vec{v} f(\vec{r},\vec{v})$ for some astrophysical system.
$f$ is understood to be some average of the ``true'' distribution $\bar{f}$, which is a sum of $\delta$-functions, each representing a star.
}
\begin{equation}
\label{eq:realization_f}
    \bar{f} \left(\vec{r},\vec{v},\{\vec{r}_i,\vec{v}_i\}\right) = \sum_{i=1}^N \delta(\vec{r}-\vec{r}_i) \delta(\vec{v}-\vec{v}_i)
\end{equation}
\REVNEW{
The ensemble average of some functional $F$ is the average over all possible ``realizations'' $\bar{f}$ of the distribution $f$ (the functional that interests us is the collision rate, see appendix~\ref{appendix:nsv_in_average}).
}
\begin{equation}
\label{eq:ensemble_avg_def}
    \left< F \right>_f = \left[ \prod_{i=1}^N \int { \frac{f(\vec{r}_i,\vec{v}_i)}{N} {\rm d} \vec{r}_i {\rm d} \vec{v}_i } \right]
    F\left[ \bar{f} \left( \vec{r},\vec{v},\left\{\vec{r}_i,\vec{v}_i\right\}_{i=1}^N \right) \right]
\end{equation}
\REVNEW{
In ergodic systems, this ensemble average is equivalent to a long-time average of any given realization;
but in our degenerate systems, this is not the case.
What we calculate in the following subsections are the long-time averaged collision rates (given a realization), and we compare them to the ensemble-averaged rates.
To simplify the discussion, we forego a full phase-space distribution description in favor of order-of-magnitude considerations.
}

\subsection{Spherically symmetric potentials}
\label{subsec:dynamics_radial}

Trajectories in a spherically symmetric potential are confined to a single plane of motion, as radial forces cannot torque an orbital angular momentum vector.
In this subsection, we examine how collisions behave under these conditions.

Here we assume that each particle's trajectory is ergodic inside its fixed plane of motion.
In other words, we will use an in-plane areal density function $g(r)$ to describe the probability of finding a particle in a thin annulus of radius $r$.
$g(r)$ is related to the volumetric density distribution $n(r)$, describing the ensemble-averaged particle density in a thin shell of radius $r$.
\REVDEL{
More quantitatively, said probability is $g(r) 2\pi r {\rm d}r = n(r) 4\pi r^2 {\rm d}r$, so that $g(r) = 2r n(r)$.
}
\REVNEW{
\begin{equation}
\label{eq:g_n_relation}
\begin{split}
    \mathbb{P}\left[ \text{radius} \in (r,r+{\rm d}r) \right] & = g(r) 2\pi r\,{\rm d}r = n(r) 4\pi r^2 {\rm d}r , \\
    g(r) & = 2r\cdot n(r).
\end{split}
\end{equation}
}

Now let us consider two particles, each with its own plane of motion.
Denoting the angle between the planes by $\alpha$, it is evident that the smaller $\alpha$ is, the higher the collision rate between the particles will be.
For $\alpha=0$, the co-planar case, we can perform a planar $\nsv$ calculation to evaluate the collision rate, with the planar cross-section $\Sigma_{2d}=2D$\REVNEW{ (where $D$ is the diameter of a hard ball with the same cross-section)}.
\begin{equation}
\begin{split}
    \text{co-planar rate}
    & = \int{d^2\vec{r} g_1 g_2 \Sigma_{2d} v} \\
    & = \Sigma_{2d} \int{2\pi r \left( 2r n(r) \right)^2 v(r) {\rm d}r} \\
    & = 16 \pi D \int{n^2 r^3 v {\rm d}r}.
\end{split}
\label{eq:coplanar_rate}
\end{equation}

Compare this to the mean ergodic rate, calculated using a volumetric $\nsv$, with $\Sigma_{3d} = \pi D^2$.
\begin{equation}
\begin{split}
    \text{mean rate} & = \int{d^3\vec{r} n_1 n_2 \Sigma_{3d} v} = \Sigma_{3d} \int{4\pi r^2 n^2 v {\rm d}r} \\
    & = 4 \pi^2 D^2 \int{n^2 r^2 v {\rm d}r}.
\end{split}
\label{eq:mean_rate_spherical}
\end{equation}

Their ratio is
\begin{equation}
    \frac{\text{co-planar rate}}{\text{mean rate}} = \frac{4}{\pi D} \frac{\int{n^2r^3v{\rm d}r}}{\int{n^2r^2v{\rm d}r}} \sim \frac{R}{D},
\label{eq:coplanar_vs_mean_spherical}
\end{equation}
where $R$ is the characteristic radius of the particles' trajectories.

\begin{figure}
    \centering
    \begin{tikzpicture}
        \def\R{2}  
        \def\a{\R} 
        \def\b{0.5*\R}  
        \def\L{0.15*\R}
        \fill[red!30] (-\R,-\L) rectangle (\R,\L);
        \draw (0,0) circle (\R);
        \draw (0,0) ellipse [x radius=\a, y radius=\b];
        \draw[dashed] (-\R,0) -- (\R,0);
        \draw[red, line width=1pt] (0,-\L) -- (0,\L);
        \node[red] at (0.5*\L,1.75*\L) {$2L$};
        \def\s{0.3*\R}  
        \def\originx{1.7*\R}
        \def\originy{-1.5*\L}
        \def\lowercenter{(\originx+3*\s, \originy)}
        \def\uppercenter{(\originx+3*\s, \originy+2*\s)}
        \draw (\originx-1.8*\s, \originy) -- (\originx+5*\s, \originy);
        \draw (\originx-1.5*\s, \originy-1*\s) -- (\originx+4.5*\s, \originy+3*\s);
        \draw[cyan, line width=0.6pt] \lowercenter circle (\s);
        \draw[cyan, line width=0.6pt] \uppercenter circle (\s);
        \draw[blue, line width=0.6pt] \lowercenter -- \uppercenter;
        \node[blue] at (\originx+3.25*\s, \originy+1.4*\s) {$D$};
        \draw[red, line width=1.2pt] (\originx, \originy) -- \lowercenter;
        \node[red] at (\originx+1.1*\s, \originy-0.6*\s) {$L=\frac{D}{\sin\alpha}$};
        \node at (\originx+1*\s, \originy+0.3*\s) {$\alpha$};
    \end{tikzpicture}
    \caption{
    On the left -- two planes intersecting (represented by the black ellipses).
    The red area is where collisions can happen.
    On the right -- a view from the side of the intersecting planes (the black lines).
    $L$ is the largest distance from the intersection where two particles with diameter $D$ can collide.
    }
    \label{fig:angle_factor}
\end{figure}
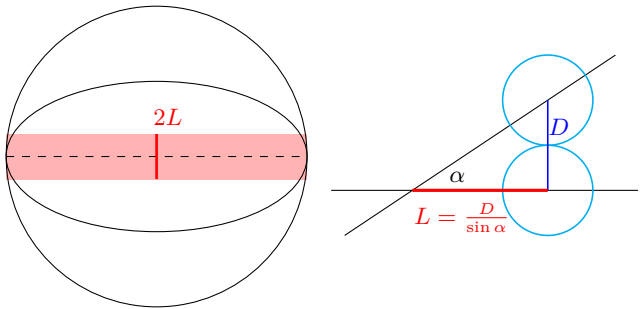

For a general angle $\alpha$, two planes intersect on a single line.
The greatest distance from this line where intersections can happen is $L=\frac{D}{\sin \alpha}$ (see figure~\ref{fig:angle_factor}).
The rate of collisions, compared to the rate where $\alpha=0$, is the ratio between the area where collisions can occur, to the total area where a particle can be.
When $L/R>1$, this ratio is close to unity; otherwise, this ratio is $\sim L/R$.
Therefore, the rate of collisions for particles at planes with relative angle $\alpha$ is:
\begin{equation}
    \text{rate}(\alpha) \sim \nsv
    \begin{cases}
        \frac{R}{D} & \text{if $\sin\alpha<\frac{D}{R}$} \\
        \frac{1}{\sin \alpha} & \text{if $\sin\alpha>\frac{D}{R}$} \\
    \end{cases}
    \label{eq:rate_spherical_by_angle}
\end{equation}

For particles with planes close to being parallel, the collision rate is greater than the ergodic rate by a factor of $\sim R/D$, but they are scarce.
The probability of the relative angle between planes being $\alpha<\frac{D}{R}$ is $\sim\frac{D^2}{2R^2}$.
Thus, the contribution of such collisions is $\sim \frac{D}{R} \ll 1$ of the total rate.

\subsection{Kepler potential: closed orbits}
\label{subsec:dynamics_kepler}

The Kepler potential is more degenerate than a general spherically symmetric potential, resulting in closed orbits.
This time we will use an even weaker ergodic hypothesis -- ergodicity along the closed orbit\footnote{Note that this hypothesis will not be valid for two closed orbits with a rational period commensurability, i.e. a mean motion resonance.}.
That is, we will assume that the appearances of particles along their trajectory can be described by a time-independent probability distribution.
This probability function is $f=1/vT$, where $v$ is the speed at a point along the orbit, and $T$ is the orbital period.

Analogously to subsection~\ref{subsec:dynamics_radial}, let us first estimate the collision rate between two particles that share the same orbit.
\begin{equation}
    \parbox{4.5em}{\centering co-orbital \\ rate}
    = \oint{d\ell f_1 f_2 \Sigma_{1d} w} \sim \frac{2\pi R w}{(vT)^2} \sim \frac{w}{v} T^{-1},
\label{eq:co_orbital_rate}
\end{equation}
where $\Sigma_{1d}=1$ is the linear ``cross-section'' (note that for perfectly closed orbits, this is just a binary $\delta$-function), $w = \left|\vec{v}_1-\vec{v}_2\right|$ is the relative speed, and $2\pi R$ is the approximate length of the orbit.
That is, if the two particles do not move in the same direction along the orbit, they will collide twice every period.

In the general case of two particles in different orbits, a collision is possible only if the two orbits intersect.
For our purposes, we define an intersection to be a local minimum of the distance between the two orbits $\Delta$, where $\Delta < D$.

The portion of a trajectory where a collision can occur is of length $2\sqrt{D^2-\Delta^2}\csc \alpha$, where $\alpha$ is the angle between the trajectories at their closest point.
Thus, the collision rate due to an intersection of distinct closed orbits is
\begin{equation}
\label{eq:collision_rate_of_intersection}
\begin{split}
    \parbox{6.5em}{\centering collision rate \\ of intersection}
    & = \int{d\ell \frac{\Sigma_{1d} w}{v_1 T_1 v_2 T_2}} \\
    & = \frac{2w\sqrt{D^2-\Delta^2}}{v_1 T_1 v_2 T_2 \sin\alpha}.
\end{split}
\end{equation}
It should be noted that for very small angles $\alpha$, this formula can predict a rate much higher than any of $T_i^{-1}$.
In such cases, equation~\ref{eq:collision_rate_of_intersection} would not be applicable, since the formula for the length of the intersection region is only valid when this region is small in comparison to the entire orbit.

A rough approximation of equation~\ref{eq:collision_rate_of_intersection} gives
\begin{equation}
\label{eq:rough_collision_rate_of_single_intersection}
    \parbox{6.5em}{\centering collision rate \\ of intersection}
    \sim \frac{D}{v T^2} \sim \frac{D}{R} T^{-1} .
\end{equation}

The expected number of intersections $\left< N_{int} \right>$ in a system with $N$ particles is roughly $\sim N^2 \frac{D}{R}$, giving an average total collision rate of
\begin{equation}
\label{eq:rough_intersection_rate}
\begin{split}
    \text{mean total rate} & = \left< N_{int} \right> \cdot \text{collision rate of intersection} \\
    & \sim \frac{N^2 D^2}{R^2 T} \sim N\nsv .
\end{split}
\end{equation}

See appendix~\ref{appendix:nsv_in_average} for a general proof that $\nsv$ is always correct in an ensemble average.
Note that the expected number of intersections may be $<1$ in some systems.
Most realizations of such systems will not have collisions at all; conversely, some realizations will have a rate of collisions far greater than $N\nsv$, with intersections yielding collisions repeatedly.

\subsection{Harmonic potential: closed orbits, universal period}
\label{subsec:dynamics_harmonic}

The harmonic potential, besides having closed orbits, has another important feature that distinguishes its behavior from the Kepler potential.
Because all trajectories have the same period, the assumption of ergodicity inside the orbit is no longer valid.
That is because even if two orbits intersect, they will either collide every period or never collide.

Given an intersection of orbits, the probability of the relative orbital phase allowing collisions is $\sim D/R$.
The expected number of such \emph{opportunities} (intersection plus phase coincidence) is $\frac{D}{R} \left< N_{int} \right> \sim N^2 \left( \frac{D}{R} \right)^2$.

\subsection{Summary of degenerate potential dynamics}
\label{subsec:dynamics_summary}

Each of the degenerate potentials discussed in this section has a special case where pairwise collision rates are enhanced.
In general spherically symmetric potentials, this is when the two particles are coplanar.
In the Kepler potential, this is when the orbits intersect.
In the harmonic potential, this case is when the orbits intersect and their relative phase permits collisions.

Table~\ref{tab:pairwise} summarizes the probability and collision rates of each case.
Multiplying the probabilities by the pairwise collision rate, we can see than regardless of the potential, the expected collision rate of a pair is $\sim \left(\frac{D}{R}\right)^2 T^{-1}$.

\begin{center}
\begin{table*}
    \begin{tabular}{|c|c|c|c|c|c|} \hline 
    \multirow{2}{*}{\textbf{Pair type}} & \multirow{2}{*}{\textbf{Probability}} & \multicolumn{4}{c|}{\textbf{Collision rate}}\\
    \cline{3-6}
    & & General potential & Spherical & Kepler & Harmonic \\
    \hline 
    
    common & $1$ & \cellcolor{green!25} $\left(\frac{D}{R}\right)^2 T^{-1}$ & \cellcolor{green!25} $\left(\frac{D}{R}\right)^2 T^{-1}$ & $0$ & $0$ \\
    \hline 
    
    co-planar & $\left(\frac{D}{R}\right)^2$ & \cellcolor{gray!25} & $\frac{D}{R} T^{-1}$ & $0$ & $0$ \\
    \hline 
    
    intersecting & $\frac{D}{R}$ & \cellcolor{gray!25} & \cellcolor{gray!25} & \cellcolor{green!25} $\frac{D}{R} T^{-1}$ & $0$ \\
    \hline 
    
    intersecting + phase & $\left(\frac{D}{R}\right)^2$ & \cellcolor{gray!25} & \cellcolor{gray!25} & \cellcolor{gray!25} & \cellcolor{green!25} $T^{-1}$ \\
    \hline
    
    \end{tabular}
    \caption{Types of particle pairs, their probability and their per-particle collision rate in different potentials.
    A common pair is the most likely case, a co-planar pair is when both particles are in the same plane (up to an angle of $D/R$), an intersecting pair is when the orbits intersect, and an intersecting + phase aligned pair is when the orbits intersect and the particles pass at the intersection at the same time.
    For each potential, the green cell is the type of pair that contributes the most to collision rate.  We note that this table focuses on idealized potentials, neglecting complications due to destructive collisions, orbital precession, etc.}
    \label{tab:pairwise}
\end{table*}
\end{center}

The essential difference between the Kepler or harmonic potentials, and the general or spherical potentials, is the type of pair the contributes most of the collisions.
In a general or a spherical potential, most collisions come from generic orbital pairings; conversely, in the Kepler or harmonic potentials, collisions come from rare types of orbit pairs, colliding repeatedly.
This is why in a general or in a spherical potential, different realizations will have roughly the same collision rate; while in the Kepler or in the harmonic potential, different realizations may have very different collision rates.

\section{The role of destructive collisions and orbital precession}
\label{sec:role_of_prec}

Up to now, we have discussed dynamical systems with perfect potentials and non-destructive (repeatable) collisions.
In this section, we will consider two more realistic effects -- destructive collisions and orbital precession.

In closed-orbit systems (Kepler and harmonic, see subsection~\ref{subsec:dynamics_summary}), collisions happen between the same particles (those with intersecting orbits) over and over again.
In light of this, the rate of collisions in such systems should be drastically diminished if collisions are {\it destructive}.

For the following discussion, to unify the treatment of the Kepler and the harmonic potential, let us call the situation where a pair of particles can collide an \emph{opportunity}.
An opportunity is an intersection for the Kepler potential, and an intersection with aligned phase for the harmonic potential.

Once the system is formed, it is only a matter of time until all opportunities yield a collision, and there will be no more collisions.
We call the characteristic time for an opportunity to yield a collision the \emph{depletion time}, $t_\text{dep}$.
The depletion time is the inverse of the collision rate for an opportunity.
\begin{equation}
\label{eq:t_dep}
    t_\text{dep} = 
    \begin{cases}
        \frac{R}{D}T & \text{Kepler} \\
        T & \text{harmonic}
    \end{cases}
\end{equation}

As a rough approximation, for a duration of $t_\text{dep}$ since the creation of the system, the rate of collisions is $\sim \frac{N_\text{op}}{t_\text{dep}}$ (where $N_\text{op}$ is the number of opportunities), and after that there are no collisions anymore.
Let us recall that the expected value of $N_\text{op}$ is
\begin{equation}
\label{eq:N_op}
    \left<N_\text{op}\right> = 
    \begin{cases}
        N^2 \frac{D}{R} & \text{Kepler} \\
        N^2 \left(\frac{D}{R}\right)^2 & \text{harmonic}
    \end{cases}.
\end{equation}

Now let us consider systems that are only approximately Keplerian or harmonic, where orbits precess with a precession time of $t_\text{prec} \gg T$.
For now, let us assume that precession alters the orbit, without changing the phase.

Precession effectively ``refreshes'' the orbits, in the sense that old opportunities may disappear as new opportunities arise.
The characteristic time for opportunities to change due to precession will be called the \emph{refresh time}, and we shall denote it by $t_\text{ref}$.
Note that $t_\text{ref}$ is smaller than the usual precession time $t_\text{prec}$, which signifies the time required for major changes in orbit to take place: $t_\text{ref} \sim \frac{D}{R} t_\text{prec}$.
This is because a displacement of order $D$ is enough to remove existing intersections and create new ones.

Once there are opportunities, they yield collisions until they are depleted after a time $t_\text{dep}$;
after a time $t_\text{ref}$, new opportunities are formed.
If $t_\text{ref} > t_\text{dep}$, the true (``refresh-limited'') collision rate will be
\begin{equation}
\label{eq:rate_refresh_limited}
    \parbox{6em}{\centering total \\ collision rate}
    = \frac{\left<N_\text{op}\right>}{t_\text{ref}} = N^2 \frac{D}{R} t_\text{ref}^{-1} \begin{cases}
        1 & \text{Kepler} \\
        \frac{D}{R} & \text{Harmonic}
    \end{cases}.
\end{equation}
This can be compared to the $\nsv$ calculation by using $\Sigma \sim D^2$, $n \sim N/R^3$ and $v \sim R/T$.
\begin{equation}
\label{eq:rate_reduction}
\begin{split}
    \parbox{6em}{\centering total \\ collision rate}
    & = N \nsv \frac{t_\text{dep}}{t_\text{ref}} \\
    & = N \nsv \frac{T}{t_\text{ref}}
    \begin{cases}
        \frac{R}{D} & \text{Kepler} \\
        1 & \text{Harmonic}
    \end{cases}.
\end{split}
\end{equation}

\subsection{Other causes for orbit shuffling}
\label{subsec:random_walk_refresh}

Precession coherently changes the orientation of orbits.
Therefore, the time required for a change of order $\delta \ll 1$ in the orbital parameters is $\delta \cdot t_\text{prec}$;
this is why $t_\text{ref} = \frac{D}{R} t_\text{prec}$.

On the other hand, other causes of orbital changes are random in nature, and behave like a random walk or a diffusion process.
The most relevant example is weak scatterings from individual particles in a many-body gravitational system.
The time for such scatterings to make an order unity change to orbital parameters is the relaxation time, $t_\text{relax}$.
The required time for a change of order $\delta$ in this case is $\delta^2 \cdot t_\text{relax}$.
Hence,
\begin{equation}
\label{eq:t_ref_t_relax}
    t_\text{ref} = \parfrac{D}{R}^2 t_\text{relax}.
\end{equation}

It should be noted that equation~\ref{eq:t_ref_t_relax} is only valid in the diffusion limit, i.e. when a single random walk ``step'' is small compared to $\delta$.
When this is not the case, $t_\text{ref}$ must be calculated according to the specifics of the random walk process, and in particular the way a random walk step affects the trajectory.
The simplest situation is where each step occurs instantaneously (compared to the trajectory time scale $T$);
then $t_\text{ref}$ will be the average time between steps.

\subsection{Imperfect harmonic systems}
\label{subsec:imperfect_harmony}

In the centers of realistic star clusters, the potential will be close to a harmonic potential, but not perfectly harmonic.
In such cases, the imperfection of the potential will cause two effects -- precession, and deviations from the universal frequency.
We have already discussed the role of precession; in this subsection, we will discuss the effect that deviations from the universal frequency have on the collision rate in nearly harmonic systems.

Let us denote the characteristic deviation from the universal frequency by $\varepsilon$, i.e. given two particles, their period ratio will be $\sim 1+\varepsilon$.
If $\varepsilon \gtrsim \frac{D}{R}$, the system is effectively not harmonic, since after one ``universal'' period a particle will no longer overlap its previous position at all.
Such a system will no longer exhibit ``harmonic'' behaviour (closed orbits plus a universal frequency), and will behave more like a ``Keplerian'' system (possessing just closed orbits):
an opportunity will be any intersection, and the depletion time will be $T\frac{R}{D}$.

On the other hand, if $\varepsilon \ll \frac{D}{R}$, there are several different cases, depending on the value of $\varepsilon \frac{t_\text{ref}}{T}$ -- the change in phase during a refresh time.
If $\varepsilon \frac{t_\text{ref}}{T} < \frac{D}{R}$, phase changes are minor in a refresh time.
In this case, variations in orbital periods are negligible and the system can be considered harmonic.

If $\frac{D}{R} < \varepsilon \frac{t_\text{ref}}{T} < 1$, not every intersection will be an opportunity, but the phase match for an opportunity is more lenient than $\frac{D}{R}$.
In this case, the expected number of opportunities is $\left< N_\text{op} \right> = N^2 \frac{D}{R} \varepsilon \frac{t_\text{ref}}{T} > \left<N_\text{op}\right>_{\text{Harmonic}}$, and the collision rate is
\begin{equation}
    \parbox{6em}{\centering total \\ collision rate}
    = \frac{\left<N_\text{op}\right>}{t_\text{ref}} = \varepsilon N^2 \frac{D}{R} T^{-1}.
\end{equation}
We call this type of behavior \emph{semi-harmonic}.

Finally, if $\varepsilon \frac{t_\text{ref}}{T} > 1$, then any intersection will yield a collision in a refresh time -- so every intersection is an opportunity, like in a Kepler potential.

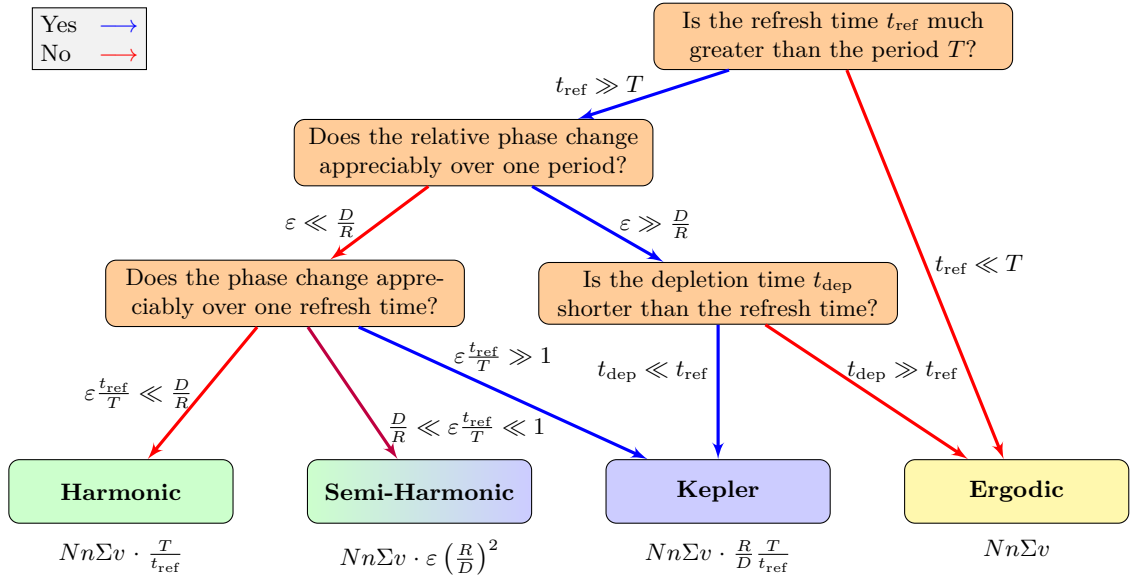
\begin{figure*}
    \centering
    \tikzset{
        decision/.style={rectangle, draw, fill=orange!40, 
            text width=14em, text centered, rounded corners},
        result/.style={
            rectangle, draw,
            text width=8.5em, minimum height=2.6em, rounded corners, align=center,
            append after command={
                \pgfextra
                    \node[below=0.3em of \tikzlastnode, text width=8em, align=center] 
                          (\tikzlastnode-desc) {#1};
                \endpgfextra
            }
        },
        yes_line/.style={draw=blue, very thick, text=black, -latex'},
        no_line/.style={draw=red, very thick, text=black, -latex'},
        hmm_line/.style={draw=purple, very thick, text=black, -latex'},
        block/.style={rectangle, draw, fill=lightgray!20, text width=4em},
    }
    \begin{tikzpicture}[node distance=2cm]
        \node [decision] (dec1) {Is the refresh time $t_\text{ref}$ much greater than the period $T$?};
        \node [decision, below left = 2em and 0em of dec1] (dec2) {Does the relative phase change appreciably over one period?};
        \node [decision, below left=3em and -7em of dec2] (dec3) {Does the phase change appreciably over one refresh time?};
        \node [decision, right=3em of dec3] (dec4) {Is the depletion time $t_\text{dep}$ shorter than the refresh time?};
        \node [result=$N\nsv\cdot\frac{R}{D}\frac{T}{t_\text{ref}}$, below=5.5em of dec4, fill=blue!20] (out3) {\textbf{Kepler}};
        \node [result=$N\nsv$, right=3em of out3, fill=yellow!40] (out4) {\textbf{Ergodic}};
        \node [result=$N\nsv\cdot\varepsilon\left(\frac{R}{D}\right)^2$, left=3em of out3, left color=green!20, right color=blue!20] (out2) {\textbf{Semi-Harmonic}};
        \node [result=$N\nsv\cdot\frac{T}{t_\text{ref}}$, left=3em of out2, fill=green!20] (out1) {\textbf{Harmonic}};
        \path [yes_line] (dec1) -- node[above left = -0.4em and 0em] {$t_\text{ref} \gg T$} (dec2);
        \path [no_line] (dec1) -- node[right] {$t_\text{ref} \ll T$} (out4);
        \path [no_line] (dec2) -- node[left = 0.5em] {$\varepsilon \ll \frac{D}{R}$} (dec3);
        \path [yes_line] (dec2) -- node[right = 0.5em] {$\varepsilon \gg \frac{D}{R}$} (dec4);
        \path [no_line] (dec3) -- node[left] {$\varepsilon\frac{t_\text{ref}}{T} \ll \frac{D}{R}$} (out1);
        \path [hmm_line] (dec3) -- node[below right = 0.4em and 1em] {$\frac{D}{R} \ll \varepsilon\frac{t_\text{ref}}{T} \ll 1$} (out2);
        \path [yes_line] (dec3) -- node[above=0.5em] {$\varepsilon\frac{t_\text{ref}}{T} \gg 1$} (out3);
        \path [yes_line] (dec4) -- node[above left] {$t_\text{dep} \ll t_\text{ref}$} (out3);
        \path [no_line] (dec4) -- node[above right = 0em and -1.2em] {$t_\text{dep} \gg t_\text{ref}$} (out4);
        \node [block, above left = 2em and 6em of dec2] (legend0) {
        Yes\hfill\textcolor{blue}{$\longrightarrow$} \\
        No\hfill\textcolor{red}{$\longrightarrow$}
        };
    \end{tikzpicture}
    \caption{
    A flow chart summarizing the collision dynamics that different systems will exhibit, and the total collision rate for each type of dense star cluster.
    }
    \label{fig:decision_tree}
\end{figure*}

Figure~\ref{fig:decision_tree} schematically summarizes the conditions for each type of behaviour.

\section{Astrophysical examples}
\label{sec:examples}

In this section, we will explore some examples of multi-particle astrophysical systems that are governed by a Kepler or a harmonic potential (to leading order).

\subsection{Nearly isothermal star clusters}
\label{subsec:clusters}

In self-gravitating systems, a spherically symmetric, uniform density distribution creates a harmonic potential.
While star clusters are not uniform density systems in general, the process of collisional relaxation will, over time, transform arbitrary initial distributions of stars into an isothermal distribution, with a constant density core surrounded by a non-uniform halo \citep{SpitzerHart71}.
Because relaxation times scale inversely with stellar densities, it is the densest star systems -- ones where $\nsv$ close encounter rates are high -- that will be most able to achieve isothermal, constant density cores over a Hubble time.
Globular clusters and nuclear star clusters (NSCs), the densest star systems in the Universe, are often approximately spherical, and possess cores with approximately uniform density;
we will parametrize core properties with a core radius $r_c$, and a central mass density $\rho_c$.
Characteristic values are shown in table~\ref{tab:clusters}.

\begin{table}
\centering
    \begin{tabular}{|c|c|c|}
        \hline
         & Globular  & Nuclear \\
         \hline
        Core radius $r_c$ & 1 pc & 2-5 pc \\
        \hline
        Total mass $M_c$ & $10^5 \, M_\odot$ & $10^6-10^8 \, M_\odot$ \\
        \hline
        Central density $\rho_c$ & $5\cdot10^3 \, M_\odot \text{pc}^{-3}$ & $10^3 - 10^5 \, M_\odot \text{pc}^{-3}$ \\
        \hline
    \end{tabular}
\caption{
Properties of globular clusters and nuclear star clusters.
The values for globular clusters are from \citet{binney_tremaine_galactic_dynamics} and the values for NSCs are from \citet{neumayer2020}.
}
\label{tab:clusters}
\end{table}

As the central potential inside an astrophysical cluster is only approximately harmonic, we will use the method outlined in figure~\ref{fig:decision_tree} to determine the collision rate.
First, we must identify sources of orbit shuffling and estimate $t_\text{ref}$.
The two main sources would be \emph{mass precession} due to the true density profile deviating from a uniform distribution, and the effect of \emph{granularity} -- gravitational encounters with individual stars in the cluster. 

\REVNEW{
We can show that $t_\text{ref} < T$ using a lower bound on the effect of granularity during one period.
Consider a test star, gravitationally interacting with all other stars in the cluster;
a star with mass $m$ and characteristic distance from the test star $r$ induces a change of velocity $\Delta v \sim \frac{Gm}{r^2}T$ over one period.
We consider a single period to avoid any assumptions on whether interactions over a longer time scale are correlated or not (see \cite{rauch1996}).
}

\REVNEW{
The greatest characteristic distance is of order $\sim r_c$, so a lower bound is $\Delta v \sim \frac{Gm}{r_c^2}T \sim \frac{m}{M_\text{tot}} v$.
The directions of the impulses from the different stars are uncorrelated, so the total velocity change over a period is
}
\begin{equation}
\label{eq:cluster_period_change}
    \frac{\Delta v_\text{tot}}{v} \sim \sqrt{N} \frac{m}{M_c} \sim N^{-1/2}.
\end{equation}

\REVNEW{
A refresh is when an orbit changes by $\frac{D}{r_c}$.
Stars, even red supergiants with $D=3\cdot10^3 R_\odot \sim 10^{-4} \text{ pc}$ \citep{levesque2005}, will always be too small for the change over a period to be less than $\frac{D}{r_c}$ (see table~\ref{tab:clusters} for values of $N$ and $r_c$).
}

\REVDEL{Removed the calculation that was implicitly based on $\nsv$.}

If instead of considering stellar collisions, we look for disruptions (or ``ionizations'') of binary pairs, the cross-section becomes much greater since $D$ will be half the separation between the stars.
An upper bound for the separation of a long-lived binary star system is given by the ``hard-soft boundary'' \citep{Heggie75, fregeau2006}: $G m_B /a_B > \sigma^2$, where $m_B$ is the binary's mass, $a_B$ is its' semi-major axis, and $\sigma^2 \sim GM_c / r_c$ is the velocity dispersion in the cluster core.
Thus, for binary disruptions,
\begin{equation}
    \frac{D}{r_c} \sim \frac{a_B}{r_c} < \frac{m_B}{M_c} = \frac{m_B}{N \bar{m}} ,
\end{equation}

The largest plausible $\frac{m_B}{\bar{m}}$ is $\sim 10^2$ (for a binary containing a large stellar mass black hole), so for any value of $N$ in the range $(10^5,10^8)$, the refresh time is still shorter than the period.

In conclusion, $\nsv$ collision rate calculations are always highly valid in star clusters, due to the effect of granularity\footnote
{We have not taken into account mass precession which is another major source of orbit shuffling, since granularity is strong enough by itself.
Mass precession would restore the $\nsv$ limit by itself, except for orbits very close to the center.}.

\subsection{Supermassive Black Holes}
\label{subsec:SMBH}
In the nearly Keplerian potential of an SMBH, orbits come close to closing, but different forms of precession play an important role in regulating collision rates, as we show here.

\subsubsection{Stellar collisions around a SMBH}

For stars orbiting a SMBH, the dominant effects that cause precession are the collective gravitational potential of all other nearby stars, and general relativistic corrections to the SMBH's nearly Keplerian potential \citep{Merritt+10}.
The precession due to other stars' gravity, which is often called \emph{mass precession}, sets an upper bound for the radius $r$ where non-ergodic behavior may occur;
conversely, precession from general relativity (GR) sets a lower bound on these radii.
We show in appendix \ref{appendix:prec_vs_relax} that, while stellar granularity/relaxation can sometimes play a role in refreshing orbital intersections, it is always subdominant to mass precession in situations of interest (unlike the situation in isothermal star clusters).

Let us assume a power-law density distribution $n \sim r^{-\alpha}$ for stars around the SMBH (in a relaxed single-species stationary state, $\alpha=7/4$; \citealt{BahcallWolf76}), normalized using a quasi-empirical formula for the influence radius $R_{\rm inf}=1\text{ pc}\sqrt{\frac{\MBH}{10^6M_\odot}}$ \citep{StoneMetzger16}.
The influence radius is defined as the radius inside of which the enclosed stellar mass $M(r)$ equals the SMBH mass $\MBH$.
Assuming that the mean stellar mass does not depend on the distance from the SMBH, this gives an enclosed-mass profile
\begin{equation}
\begin{split}
\label{eq:smbh_mass_profile}
    M(r) =& \MBH \parfrac{r}{R_{\rm inf}}^{3-\alpha} \\
    =& \MBH \mmm^{\frac{\alpha-3}{2}} \inpc{r} ^{3-\alpha}.
\end{split}
\end{equation}

The mass precession time inside the radius of influence is $t_\text{prec}^{\text{mass}} \sim \frac{\MBH}{M(r)} T$, which leads to a refresh time of
\begin{equation}
    t_\text{ref}^{\text{mass}} \sim \frac{D}{r} \frac{\MBH}{M(r)} T,
\end{equation}
and a depletion/refresh time ratio of
\begin{equation}
\label{eq:ratio_mass_SMBH}
\begin{split}
    & \frac{t_\text{dep}}{t_\text{ref}^{\text{mass}}} 
    = \parfrac{r}{D}^2 \frac{M(r)}{\MBH} \\
    & = 5\cdot10^{14} \parfrac{D}{2R_\odot} ^{-2} \mmm^{\frac{\alpha-3}{2}} \inpc{r} ^{5-\alpha}.
\end{split}
\end{equation}
In contrast, the GR precession time is $\sim \frac{r}{R_{\rm BH}} T$, where $R_{\rm BH} = \frac{2G\MBH}{c^2}$ is the Schwarzschild radius of the black hole.
So, the refresh time is
\begin{equation}
    t_\text{ref}^{\text{GR}} \sim 0.5 \parfrac{D}{2R_\odot} \mmm^{-1} T.
\end{equation}
The depletion/refresh time ratio is thus
\begin{equation}
\label{eq:ratio_GR_SMBH}
    \frac{t_\text{dep}}{t_\text{ref}^{\text{GR}}} = 2 \parfrac{D}{2R_\odot}^{-1} \frac{r}{D} \mmm .
\end{equation}

\begin{figure*}
\centering
\includegraphics[width=\textwidth]{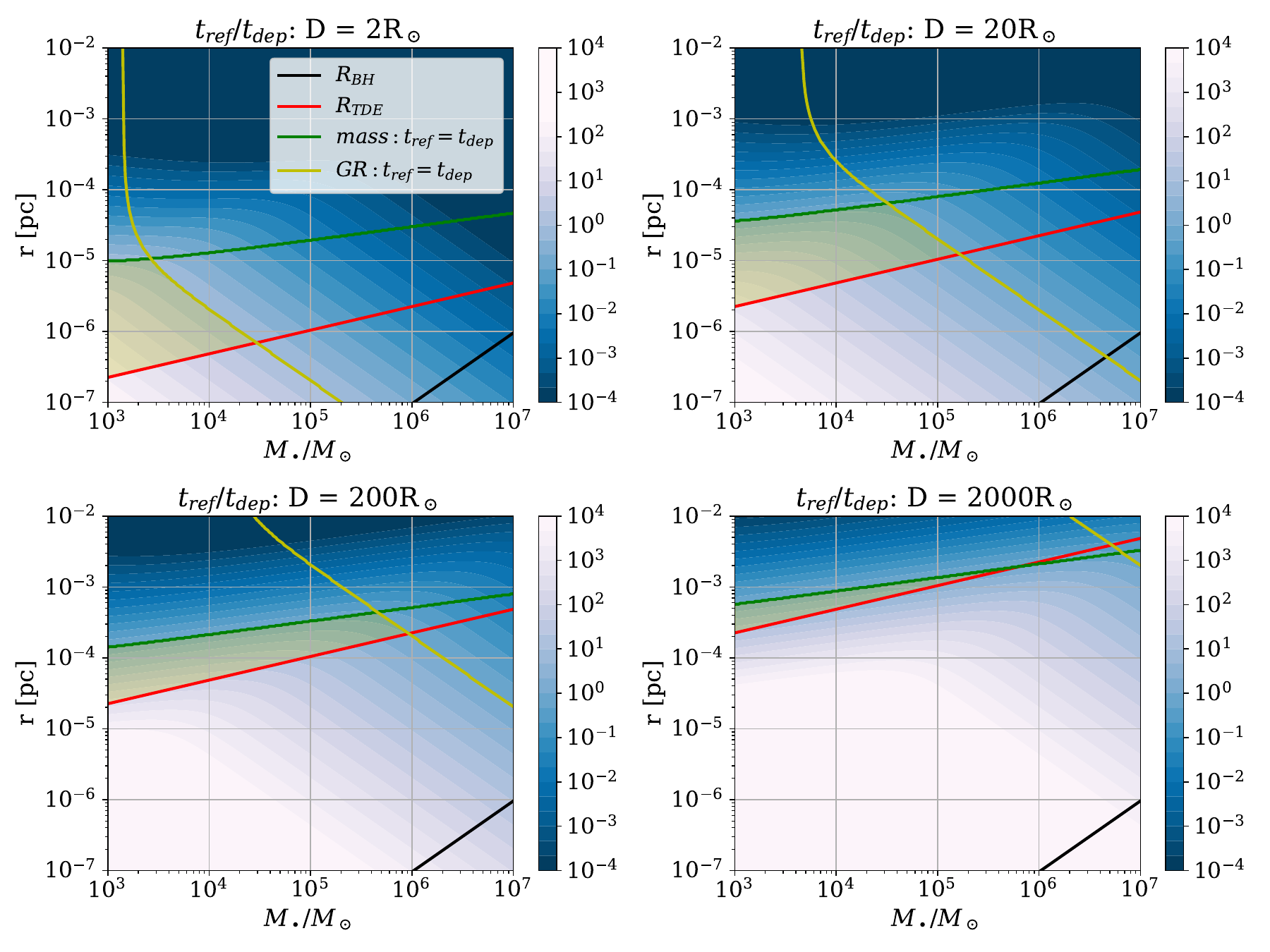}
    \caption{Color map of the ratio $t_\text{ref}/t_\text{dep}$ as a function of SMBH mass $\MBH$ and orbital radius $r$, for a star with mass $M_\odot$, and different collisional cross-sections (diameter from $2R_\odot$ to $2\cdot10^3 R_\odot$).
    The ratio is taken as the sum of reciprocals of equation~\ref{eq:ratio_mass_SMBH} and equation~\ref{eq:ratio_GR_SMBH}.
    The black line is the Schwarzschild radius of the SMBH and the red line is the tidal disruption radius of a star with diameter $D$.
    The green line and the yellow line are the lines of equal $t_\text{dep}$ and $t_\text{ref}$ for mass and GR precessions (respectively).
    The light-yellow shaded region is where the ratio is greater than unity.}
\label{fig:t_ratio_SMBH_const_d}
\end{figure*}

From equation~\ref{eq:ratio_GR_SMBH}, it is evident that,for Sun-like stars around SMBHs, the depletion time will always be much greater than the GR refresh time.
Red supergiants on the other hand, can have a diameter up to 3 orders of magnitude greater than $R_\odot$.
Figure~\ref{fig:t_ratio_SMBH_const_d} shows the ratio between depletion time and refresh time for different values of $D$, as functions of SMBH mass and distance from it.
Relevant radii must be greater than the Schwarzschild radius $\frac{2G\MBH}{c^2}$ and the tidal disruption radius $\frac{D}{2} \parfrac{\MBH}{M_\star}^{\frac{1}{3}}$, otherwise the stars will not survive long enough for collisions to be important.

The results shown in figure~\ref{fig:t_ratio_SMBH_const_d} also take into account gravitational focusing, which increases the effective collisional diameter of the star according to
\begin{equation}
\begin{split}
    D_{eff}  & = D \cdot \sqrt{1 + \parfrac{v_{\rm esc}}{v_{\rm rel}} ^2} \\
    & \approx D \sqrt{ 1 + 10^2 \frac{ \parfrac{M_\star}{M_\odot} \inpc{r} } { \mmm \parfrac{D}{2R_\bullet} } },
\end{split}
\end{equation}
where $v_{\rm esc} = \sqrt{4GM_\star/D}$ is the escape velocity from the surface of the star, and $v_{\rm rel} \approx \sqrt{G\MBH/r}$ is the relative velocity of colliding stars.

Under these constraints, there are small regions of parameter space where $t_\text{dep} < t_\text{ref}$.
In general, such regions will exist for {\it intermediate mass black holes} with masses $\MBH < 10^6 M_\odot$, and radii not much larger than the tidal disruption radius.
An extreme example in figure~\ref{fig:t_ratio_SMBH_const_d} is of $\MBH = 10^4 M_\odot$, $D = 200 R_\odot$, and $r = 5\cdot10^{-5} \text{ pc} = 10 \text{ AU}$, where the ratio $t_\text{ref}/t_\text{dep}$ is as large as $\sim 10^2$.
By equation \ref{eq:rate_reduction}, this amounts to a reduction of collision rates by a factor of $10^{-2}$ in this region.

Note that a common assumption in each panel of figure~\ref{fig:t_ratio_SMBH_const_d} is that the star's mass $M_\star = 1 M_\odot$;
the star's mass $M_\star$ is only relevant for the tidal disruption radius, so for more massive stars, there is a bigger part of parameter space where $t_\text{dep} < t_\text{ref}$.
Likewise, we assume orbits have a single characteristic radius for simplicity.

\subsubsection{Binary disruptions around a SMBH}

The interaction cross-section for collisional ionization of a binary can be much greater than that for direct physical collisions with a star.
In this case, $D$ would be, roughly, the separation between the stars in the binary system.
The possibility of the SMBH tidally disrupting the binary sets an upper bound for the possible separation in a given radius
\begin{equation}
\label{eq:dmax}
    D_{\rm max} = r \parfrac{m_{\rm B}}{\MBH}^{\frac{1}{3}} ,
\end{equation}
where $m_{\rm B}$ is the mass of the binary.

\begin{figure}
\centering
\includegraphics[width=\columnwidth]{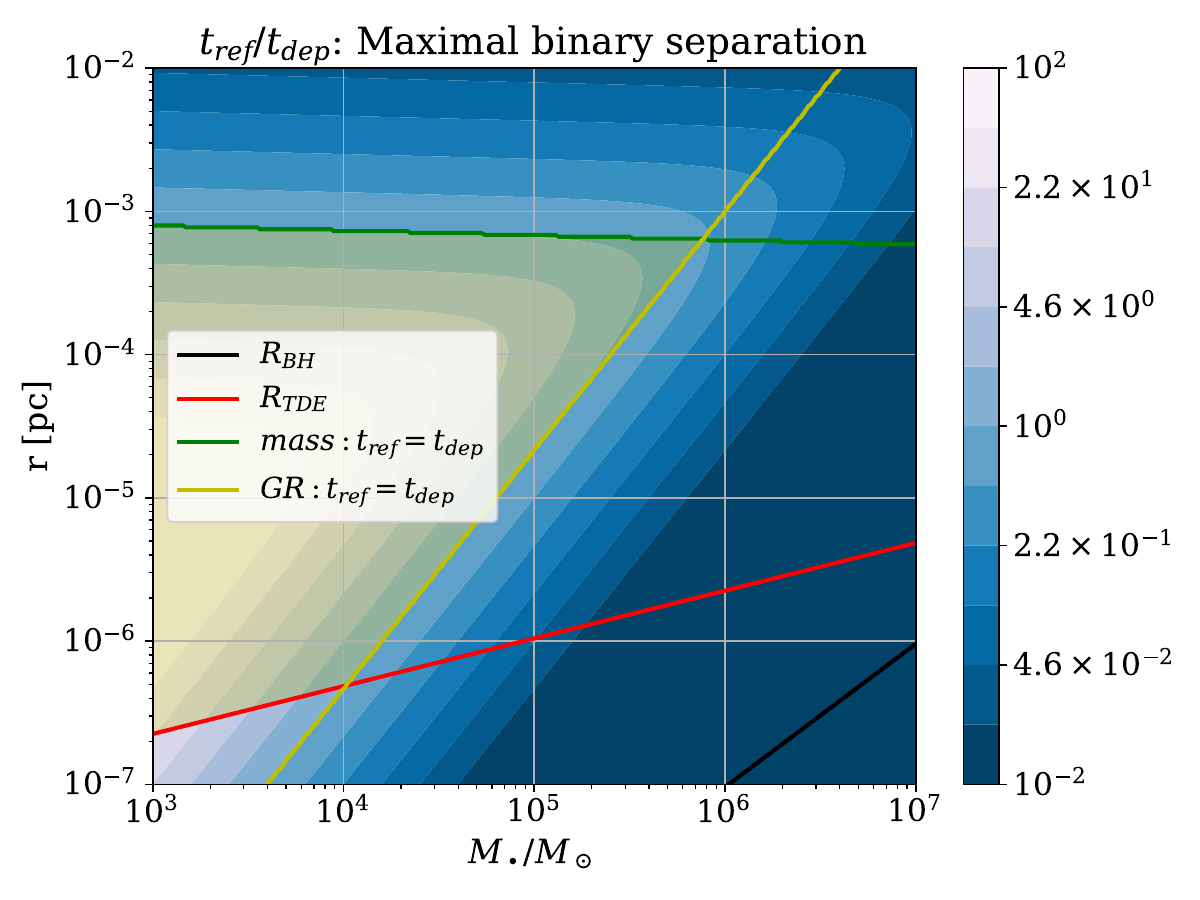}
    \caption{Color map of the ratio $t_\text{ref}/t_\text{dep}$ as a function of SMBH mass $\MBH$ and radius $r$, for collisional ionization of a binary with maximal separation (as given by equation~\ref{eq:dmax}).
    The ratio is taken as the sum of reciprocals of equation~\ref{eq:ratio_mass_SMBH} and equation~\ref{eq:ratio_GR_SMBH}.
    The black line is the Schwarzschild radius of the SMBH and the red line is the tidal disruption radius of a Sun-like star.
    The green line and the yellow line are the lines of equal $t_\text{dep}$ and $t_\text{ref}$ for mass and GR precessions (respectively).
    The light-yellow shaded region is where the ratio is greater than unity.
    Binary ionization rates can be 1-2 orders of magnitude lower than the $\nsv$ limit inside the influence radius of intermediate mass black holes.}
\label{fig:t_ratio_SMBH_binary_dmax}
\end{figure}

Figure~\ref{fig:t_ratio_SMBH_binary_dmax} shows the ratio between refresh time and depletion time for binaries with the maximal separation (by equation~\ref{eq:dmax}).
It can be seen that for black hole masses of less than $10^6 M_\odot$, there is a spatial region where refresh times can be greater than the depletion time, and non-ergodic collisional behavior may occur.
Intermediate mass black holes with $M_\bullet \lesssim 10^5 M_\odot$ exhibit a significant range of radii where binaries will experience collisional ionization at rates 1-2 orders of magnitude below the $\nsv$ prediction.
For significantly more massive SMBHs, the ergodic calculation of collision rates will always be appropriate.

\subsection{Planetesimals around a massive star}

Planetesimals in asteroid belts or in debris disks around a star move in a nearly Keplerian potential.
Over long timescales, the size distribution of these planetesimals will evolve in a collisional cascade, eventually reaching a steady state in mass flux due to collisional fragmentation \citep{Dohnanyi69}.
The detailed shape of this steady state particle size distribution is a function of the collision rate, which is generally computed in the $\nsv$ way \citep{Tanaka+96, PanSari05, Schlichting+13}.
It is therefore interesting to understand when the standard approach can be applied.

In the asteroid belt of a planetary system, the main sources of precession are GR and perturbations from a massive planet, if one exists.
In a tightly bound debris disk around a star, the main sources of precession are GR, the disk's gravitation, and the star's quadrupole moment.
In this section, we explore the conditions under which the $\nsv$ treatment of collision rates can be applied to collisional cascades in various debris disks.

\subsubsection{Asteroid belt in a planetary system}
Taking equation~\ref{eq:ratio_GR_SMBH} for GR precession, and modifying its normalizations to be more suitable for the case of an asteroid belt in a planetary system, we get
\begin{equation}
\label{eq:ratio_GR_belt}
    \frac{t_\text{dep}}{t_\text{ref}^{\text{GR}}} = 4\cdot10^8 \parfrac{D}{\text{1 km}}^{-2} \parfrac{r}{\text{1 AU}} \parfrac{M_\star}{M_\odot} ,
\end{equation}
where $M_\star$ is the mass of the central star.
If the central star has mass similar to our sun, and the orbit radius is of about 1 AU, this ratio would be less than 1 only for planetesimals with a diameter $D$ of at least $10^4\text{ km}$, comparable to the Earth.

An important mechanism that could lengthen the refresh time is low orbital eccentricity $e$, which is in any case characteristic of many objects orbiting our solar system.
If precession occurs in the plane of the orbit, its effect is to rotate the ellipse of the trajectory.
The effect of this rotation will be slight for a low-$e$ ellipse closely resembling a circle.
The refresh time is the time it takes precession to change the orbit by $D$;
for an orbit with eccentricity $e$, this time is
\begin{equation}
    t_\text{ref} = e^{-1} t_\text{ref,1},
\end{equation}
where $t_\text{ref,1}$ is the original estimate of the refresh time, $t_\text{prec} D/r$.

Another source of precession in asteroid belts is the perturbative gravitational pull of planets in the system.
Let us assume most of this effect comes from the largest planet in the system, which we will call ``Qupiter''.
We denote Qupiter's mass by $m_\text{Q}$ and the semi-major axis of its trajectory $a_\text{Q}$.
In the secular approximation, the precession time due to Qupiter is \citep{mustill_wyatt2009}
\begin{equation}
    t_\text{prec}^\text{Q} = T \parfrac{M_\text{Q}}{M_\star}^{-1} \cdot \max \left\{ \parfrac{a_\text{Q}}{r}^{-2}, \parfrac{a_\text{Q}}{r}^3 \right\}.
\end{equation}
The ratio of depletion time to refresh time is then
\begin{equation}
\label{eq:ratio_qupiter}
\begin{split}
    \frac{t_\text{dep}}{t_\text{ref}^\text{Q}} = & e \parfrac{M_\text{Q}}{M_\star} \parfrac{D}{r}^{-2} \\
    & \cdot \min \left\{ \parfrac{a_\text{Q}}{r}^{2}, \parfrac{a_\text{Q}}{r}^{-3} \right\}.
\end{split}
\end{equation}

We can use equations~\ref{eq:ratio_GR_belt} and~\ref{eq:ratio_qupiter} to determine $D_{c}$, the characteristic diameter above which we expect the collision rate to be less than the ergodic $\nsv$ estimate.
When secular precession from Qupiter dominates,
\begin{equation}
\label{eq:dc_qupiter}
\begin{split}
    D_c^\text{Q} = & 1.4\cdot10^8\text{ km} \cdot e^{1/2} \parfrac{M_\text{Q}}{M_\star}^{\frac{1}{2}} \\
    & \cdot \min \left\{ \parfrac{a_\text{Q}}{\text{1 AU}}, \parfrac{a_\text{Q}}{\text{1 AU}}^{-\frac{3}{2}} \parfrac{r}{\text{1 AU}}^{\frac{5}{2}} \right\}.
\end{split}
\end{equation}
While the effect of secular precession can always be dialed down by considering systems whose planets are lower in mass or more distant from the planetesimal belt, GR precession sets an unavoidable floor on $D_c$.
In cases where the refresh time is set by GR precession, from equation~\ref{eq:ratio_GR_belt}, $D_c$ will be
\begin{equation}
\label{eq:dc_GR}
\begin{split}
    D_c^\text{GR} & = 2\cdot10^4\text{ km} \cdot e^{1/2} \parfrac{r}{\text{1 AU}}^{\frac{1}{2}} \parfrac{M_\star}{M_\odot}^{\frac{1}{2}} \\
    & = 10^3\text{ km} \cdot \parfrac{e}{0.03}^{\frac{1}{2}} \parfrac{r}{\text{1 AU}}^{\frac{1}{2}} \parfrac{M_\star}{0.1 M_\odot}^{\frac{1}{2}}.
\end{split}
\end{equation}

It is clear from equation~\ref{eq:dc_GR} that for planetesimal belts as distant as the Kuiper belt (i.e. located at $r>30$ AU), around a star with mass $\gtrsim M_\odot$, the critical diameter will be greater than Earth's diameter.
That is true even taking into account the low eccentricity of the Kuiper belt's ``cold'' population, for which $e\sim0.03$ is representative \citep{Petit:2011wj}.
Asteroid belts with a smaller orbital radius, such as the asteroid belt in our own Solar system at $\sim2$ AU, will generally have a smaller $D_c$.
Even here, however, it is hard to get below $D_c \sim 10^{2-3}$ km unless one invokes a very dynamically cold planetesimal population.

\begin{figure}
\centering
\includegraphics[width=\columnwidth]{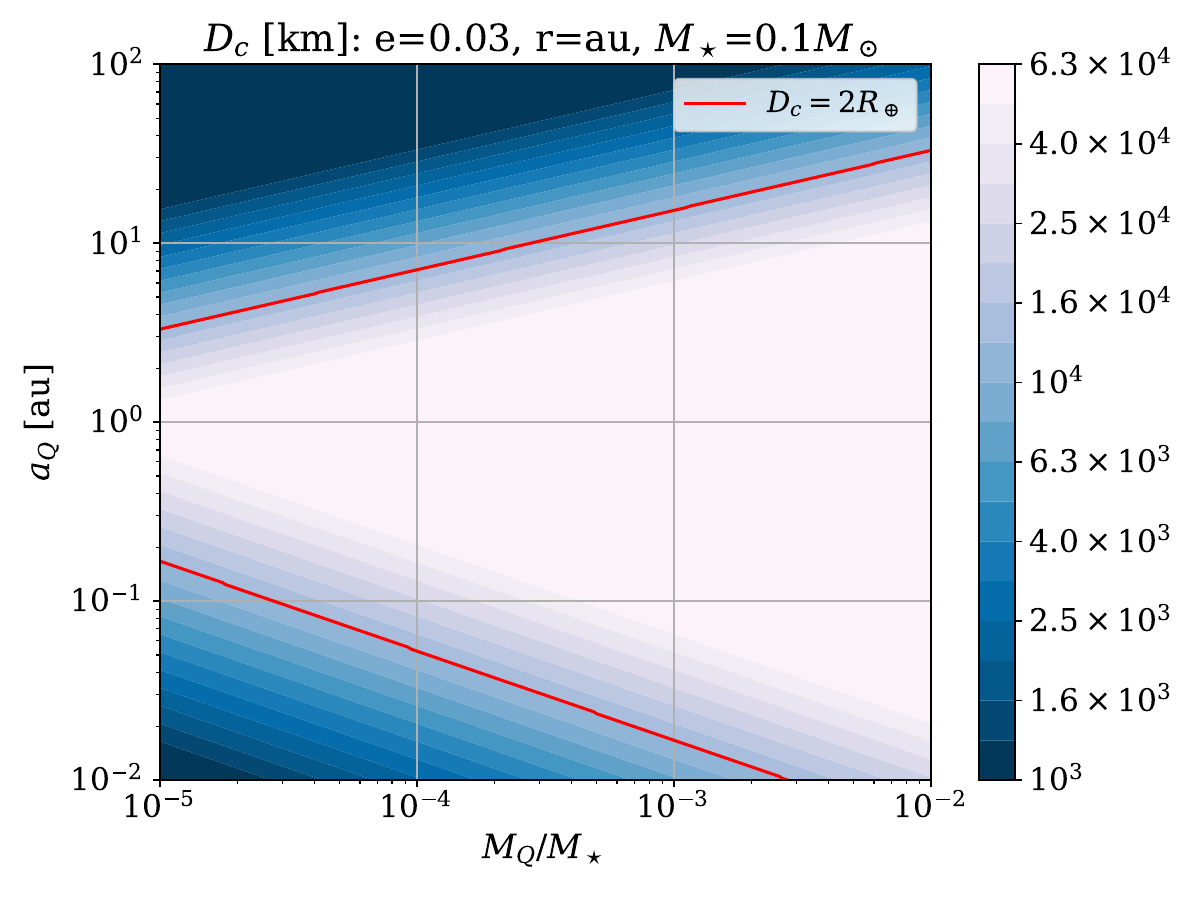}
\caption{
    Color map of $D_c$ as a function of the perturbing planet's (Qupiter's) mass and semimajor axis.
    $D_c$ is calculated according to equations~\ref{eq:dc_qupiter} and~\ref{eq:dc_GR}.
    The orbital eccentricity of planetesimals is taken to be 0.03, their orbital radius 1 AU, and the central star's mass $M_\star = 0.1 M_\odot$.
    The red line indicates where $D_c$ equals Earth's diameter.
    }
\label{fig:dc_qupiter_GR}
\end{figure}

Figure~\ref{fig:dc_qupiter_GR} shows $D_c$ as a function of a perturbing planet's (Qupiter's) mass and semimajor axis, for a planetesimal belt at 1 AU with $e=0.03$, and a central red dwarf with $M_\star = 0.1 M_\odot$.
It can be seen that if Qupiter's mass is less than about $10^{-3} M_\star = 0.1 M_\text{Jupiter}$ and its orbital radius is $\gtrsim 10$ AU, objects as large as the Earth can experience a reduced collision rate.
Deviations from ergodicity can extend to dwarf planet (e.g. Ceres) sizes only if no gas/ice giants are present, and furthermore the planetesimal belt is very low-$e$ (with $e \ll 10^{-2}$).

\subsubsection{Debris disks around white dwarves}
The tidal disruption of asteroids and dwarf planets can produce very compact debris disks around white dwarf stars \citep{Jura03, Metzger+12}.
The aftermath of such a planetesimal disruptions is observed as an infrared excess from the resulting dusty debris \citep{ZuckermanBecklin87, Debes+11}, and current work estimates that a few percent of all white dwarves have this type of debris disk at any point in time \citep{Farihi+09, Bonsor+17}.

The main sources of precession for debris orbiting a white dwarf are GR, the bulk gravitation of the debris disk itself, and the quadrupole moment of the white dwarf (from e.g. rotational oblateness).
Let us take the white dwarf and the debris disk from \citet{manser2019} as an example, with the debris disk orbiting at $r \sim 5\cdot10^{-3} \text{ AU}$, and $M_\star = 0.7 M_\odot$.
Using equation~\ref{eq:dc_GR} for the critical diameter due to GR precession, and assuming $e\approx 0.01$, we get $D_c^\text{GR} \approx 100 \text{ km}$.

Mass precession due to the debris disk's gravitation gives $t_\text{ref} \sim \frac{D}{r} \frac{M_\star}{M_\text{disk}} T$.
The critical radius due to mass precession is then
\begin{equation}
    D_c^\text{mass} = \sqrt{e \frac{M_\text{disk}}{M_\star}} r .
\end{equation}

For $D_c^\text{mass}$ to be less than $D_c^\text{GR}$,
\begin{equation}
    M_\text{disk} < 3 M_\odot \parfrac{M_\star}{M_\odot}^2 \parfrac{r}{1 \text{km}}^{-1} .
\end{equation}
In our case, this amounts to $M_{\rm disk} < 2\cdot10^{-6} M_\odot = 0.6 M_\oplus$.
That is, the critical radius will still be $\sim 100 \text{ km}$ if the total mass of the debris disk is less than about Earth's mass.

Lastly, precession due to the white dwarf's quadrupole moment $J_2$ is \citep{will_1993}
\begin{equation}
    t_\text{prec} \sim J_2^{-1} T \parfrac{r}{R_\star} ^2.
\end{equation}
The quadrupole moment can be estimated $J_2 \sim \parfrac{\omega}{\omega_c} ^2$, with the critical rotation frequency $\omega_c = \sqrt{GM_\star/R_\star^3}$.
Equating $t_\text{ref} = e^{-1} \frac{D}{r} t_\text{prec}$ and $t_\text{dep} = \frac{r}{D} T$, we get the critical size
\begin{equation}
    D_c^\text{quad} = \sqrt{e} R_\star \frac{\omega}{\omega_c} .
\end{equation}

In our case, $R_\star = 7000 \text{ km}$ and $M_\star = 0.7 M_\odot$, so $\omega_c = 7\cdot10^3 \frac{2\pi}{\text{day}}$.
To have $D_c^\text{quad} < 100 \text{ km}$, the white dwarf must rotate slower than $10^3$ times per day.
As can be seen in \citet{hermes2017}, a more likely rotation period for a white dwarf is of the order of a few rotations per day;
therefore, the WD quadrupole moment is likely a negligible contribution to precession.

To conclude, in debris disks with less mass than $\sim M_\oplus$ around slowly rotating white dwarves, it is possible for planetesimals with a diameter greater than $\sim 10^2 \text{ km}$ to experience a reduced, non-ergodic collision rate.

\section{Conclusion}

We have studied the collision rate in systems governed by degenerate potentials such as the Keplerian one and the harmonic oscillator.
Although these perfect potentials represent idealizations of any astrophysical system, they are often quite good approximations of the densest star systems in the Universe, where interesting phenomena can arise from collisions or other close encounters between stars.
Likewise, the Kepler potential is an excellent approximation for planetary and exoplanetary dynamics, where close encounter rates are of interest for e.g. understanding outcomes of collisional cascades.

We have shown that for individual realizations of such degenerate systems, the kinetic collision rate $\nsv$ is not always valid.
If collisions are non-destructive, an ensemble average over all possible realizations of a system will recover the $\nsv$ rate, but specific realizations may differ greatly from it.
In a spherical potential, $\nsv$ will be essentially correct for most specific realizations.
For the closed orbits of the Kepler potential and the harmonic potential, most realizations will not have any collisions at all, while a few realizations will have a collision rate far higher than $\nsv$.

In more realistic systems, collisions are destructive and orbits will never be perfectly closed, due to the effects of precession (i.e. bulk deviation from an idealized degenerate potential) and relaxation (i.e. deviations from the idealized degenerate potential sourced by small-scale, stochastic granularity).
We have shown how to take these effects into account, and have categorized types of systems according to the formula required to calculate the collision rate;
in an increasingly degenerate order, the categorization is: ergodic, Keplerian, semi-harmonic, and harmonic (see figure \ref{fig:decision_tree}).
The more degenerate a system is, the lower the rate of destructive collisions will be compared to $\nsv$.

While these results suggest a potential failure of the usual $\nsv$ formalism in the astrophysical contexts where collisions are of greatest interest, we have found that different physical effects will ``save'' the ergodic $\nsv$ rate in almost all collisional environments.
In isothermal star clusters (globulars and NSCs) lacking a massive central black hole, relaxation from two-body scatterings is the key physical effect that refreshes opportunities for pairwise collisions and recovers the $\nsv$ rate; we find that deviations from ergodicity in these dense star clusters are wholly negligible.
In the deeper potential wells of massive black holes, relaxation can be less efficient, and orbital intersections are generally refreshed by coherent precession (either from GR or from the extended mass of the star cluster around the black hole).
While minor deviations from the $\nsv$ collision rate can exist for some star-star collisions (figure \ref{fig:t_ratio_SMBH_const_d}, the most dramatic failure of ergodicity arises for binary ionizations inside the influence radius of intermediate mass black holes (figure \ref{fig:t_ratio_SMBH_binary_dmax}).  

In debris disks and planetesimal belts, collisional cascades can in principle occur at rates below the typical $\nsv$ one, although precession from GR as well as secular torques (from any large exoplanets in the star system) will restore the $\nsv$ limit for most objects below the size of a dwarf planet.
The impact of precession will be muted, however, for very dynamically cold planetesimal/debris disks.

\REVNEW{
In general, the smaller $D/R$ is, the more likely it is that $\nsv$ will be valid.
However, the threshold for when $D/R$ is \emph{small enough} may be several orders of magnitude below unity, e.g. in Kepler potentials with precession driving orbital changes -- the threshold is $\sqrt{\frac{T}{t_\text{prec}}}$.
}

At a high level of abstraction, there is no {\it a priori} reason why the $\nsv$ formula should apply in nearly Keplerian or nearly harmonic potentials.
The $\nsv$ approach to collision rates assumes a uniform sea of targets, but in reality, configurations of particles orbiting in nearly degenerate potentials will usually, at any moment in time, lack {\it any} opportunities for particle-particle collision.
We have shown that in different astrophysical examples of nearly degenerate potentials, collisional dynamics recovers the $\nsv$ limit due to varied combinations of GR precession, secular precession, and two-body relaxation.
There is no universal pattern as to which effect dominates, and different astrophysical systems must be evaluated on a case-by-case basis to determine why orbits fail to close sufficiently.
However, in all the examples we have considered, nearly degenerate potentials are almost never degenerate enough to avoid the classic $\nsv$ collision rate formula, and so we conclude that $\nsv$ {\it is an unreasonably effective} description of encounters in astrophysical environments.

\section*{Data availability}
The data that support the findings of this study are available within the article.
The python scripts that created the figures in this article are available at \url{https://github.com/elishamod/Non-ergodic-collision-rates}.

\begin{acknowledgments}
This research was partially supported by an ISF, MOS and an NSF/BSF grants.
NCS gratefully acknowledges support from the Israel Science Foundation (Individual Research Grant 2565/19) and the Binational Science Foundation (grant Nos. 2019772 and 2020397).
EM gratefully acknowledges support from the Milner Foundation.
The authors thank the anonymous referee for helpful and insightful comments.
\end{acknowledgments}



\appendix

\section{Equivalence to \texorpdfstring{$\nsv$}{n-Sigma-v} in an ensemble average}
\label{appendix:nsv_in_average}

A system with closed orbits and non-destructive collisions, may sometimes have a much greater collision rate than the ergodic rate $\nsv$, and sometimes have no collisions at all.
In this appendix we show that in an ensemble average, \REVDEL{which is an average over all possible realizations of a distribution,} the rate of collisions is precisely equal to the ergodic rate \REVNEW{$\nsv$}.

\REVDEL{From here on, this appendix is completely rewritten.}

\REVNEW{
Let us define the functional $R$, the instantaneous local collision rate for the phase-space distribution $f(\vec{r},\vec{v})$ at $\vec{r}$.
\begin{equation}
\label{eq:general_rate_functional}
    R(f) = \frac{1}{2} \int {\rm d}\vec{v}_1 {\rm d}\vec{v}_2 \,f(\vec{r},\vec{v}_1) \left| \vec{v}_1 - \vec{v}_2 \right| \int_\Sigma {\rm d}A \, f(\vec{r}+\vec{a},\vec{v}_2)
\end{equation}
where the $\frac{1}{2}$ factor is to account for double counting.
The $\Sigma$-integral is over the cross-section of particle 1, defined as an area of $\Sigma$ around particle 1, and perpendicular to $\vec{v}_1-\vec{v}_2$.
This way, a collision is counted at the moment of closest approach (when $\Delta\vec{r} \perp \Delta\vec{v}$).
}

\REVNEW{
First, we prove that the collision rate in an ensemble average is the same as the collision rate for the distribution function itself, i.e. $\left< R \right>_f = R(f)$, where $\left< \cdot \right>_f$ is defined in equation~\ref{eq:ensemble_avg_def}.
}

\begin{proof}

\REVNEW{
By definition of the ensemble average,
\begin{equation}
\begin{split}
    \left< R \right>_f = & \left[ \prod_{i=1}^N \int { \frac{f(\vec{r}_i,\vec{v}_i)}{N} {\rm d} \vec{r}_i {\rm d} \vec{v}_i } \cdot \right]
    R\left( \sum_{i=1}^N \delta(\vec{r}-\vec{r}_i) \delta(\vec{v}-\vec{v}_i) \right) \\
    = & \prod_{k=1}^N \int { \frac{f(\vec{r}_k,\vec{v}_k)}{N} {\rm d} \vec{r}_k {\rm d} \vec{v}_k \cdot} \\
    & \frac{1}{2} \int {\rm d}\vec{v} {\rm d}\vec{v}' \, \sum_{i=1}^N \delta(\vec{r}-\vec{r}_i)\delta(\vec{v}-\vec{v}_i) \left| \vec{v} - \vec{v}' \right| \int_\Sigma {\rm d}A \, \sum_{j=1}^N \delta(\vec{r}+\vec{a}-\vec{r}_j)\delta(\vec{v}'-\vec{v}_j) \\
    = & \frac{1}{2} \sum_{i,j} \left[ \prod_k \int { \frac{f(\vec{r}_k,\vec{v}_k)}{N} {\rm d} \vec{r}_k {\rm d} \vec{v}_k \cdot} \right]
    \delta(\vec{r}-\vec{r}_i)\left| \vec{v}_i - \vec{v}_j \right| \int_\Sigma {\rm d}A \, \delta(\vec{r}+\vec{a}-\vec{r}_j)
\end{split}
\end{equation}
The only non-trivial $(\vec{r}_k,\vec{v}_k)$ integrals are on $i$ and $j$,
\begin{equation}
    \left< R \right>_f = \frac{1}{2N^2} \sum_{i,j} \int {\rm d} \vec{r}_i {\rm d} \vec{v}_i {\rm d} \vec{r}_j {\rm d} \vec{v}_j f(\vec{r}_i,\vec{v}_i) f(\vec{r}_j,\vec{v}_j)
    \delta(\vec{r}-\vec{r}_i)\left| \vec{v}_i - \vec{v}_j \right| \int_\Sigma {\rm d}A \, \delta(\vec{r}+\vec{a}-\vec{r}_j)
\end{equation}
There are $N(N-1)$ pairs of $i \ne j$, and they all give the same contribution to the sum.
There are $N$ pairs of $i = j$, and their contribution is $0$ (due to the $\left| \vec{v}_i - \vec{v}_j \right|$ factor).
\begin{equation}
\begin{split}
    \left< R \right>_f &= \frac{N-1}{2N} \int {\rm d} \vec{r}_1 {\rm d} \vec{v}_1 {\rm d} \vec{r}_2 {\rm d} \vec{v}_2 \, f(\vec{r}_1,\vec{v}_1) f(\vec{r}_2,\vec{v}_2)
    \delta(\vec{r}-\vec{r}_1)\left| \vec{v}_1 - \vec{v}_2 \right| \int_\Sigma {\rm d}A \, \delta(\vec{r}+\vec{a}-\vec{r}_2) \\
    &= \frac{N-1}{2N} \int {\rm d} \vec{r}_2 {\rm d} \vec{v}_1 {\rm d} \vec{v}_2 \, f(\vec{r},\vec{v}_1) f(\vec{r}_2,\vec{v}_2)
    \left| \vec{v}_1 - \vec{v}_2 \right| \int_\Sigma {\rm d}A \, \delta(\vec{r}+\vec{a}-\vec{r}_2) \\
    &= \frac{N-1}{2N} \int {\rm d} \vec{v}_1 {\rm d} \vec{v}_2 \, f(\vec{r},\vec{v}_1)
    \left| \vec{v}_1 - \vec{v}_2 \right| \int_\Sigma {\rm d}A \int {\rm d} \vec{r}_2 \delta(\vec{r}+\vec{a}-\vec{r}_2) f(\vec{r}_2,\vec{v}_2) \\
    &= \frac{N-1}{2N} \int {\rm d} \vec{v}_1 {\rm d} \vec{v}_2 \, f(\vec{r},\vec{v}_1)
    \left| \vec{v}_1 - \vec{v}_2 \right| \int_\Sigma {\rm d}A \, f(\vec{r}+\vec{a},\vec{v}_2) \\
    & \xrightarrow{N\to\infty} \frac{1}{2} \int {\rm d}\vec{v}_1 {\rm d}\vec{v}_2 \,f(\vec{r},\vec{v}_1) \left| \vec{v}_1 - \vec{v}_2 \right| \int_\Sigma {\rm d}A \, f(\vec{r}+\vec{a},\vec{v}_2)
\end{split}
\end{equation}
That is precisely the functional in equation~\ref{eq:general_rate_functional}.
}

\end{proof}

\REVNEW{
To finish, let us show explicitly that \ref{eq:general_rate_functional} is in fact $\nsv$.
If we assume that $f$ does not change on the scale of distances $\sim\sqrt{\Sigma}$, then the inner integral can be simplified $\int_\Sigma {\rm d}A f(\vec{r}+\vec{a},\vec{v}_2) = \Sigma \cdot f(\vec{r},\vec{v}_2)$.
\begin{equation}
\label{eq:naive_rate_functional}
    R(f) = \frac{1}{2} \int {\rm d}\vec{v}_1 {\rm d}\vec{v}_2 \, f(\vec{r},\vec{v}_1)  f(\vec{r},\vec{v}_2) \Sigma \left| \vec{v}_1 - \vec{v}_2 \right|.
\end{equation}
If we define the mean relative velocity $v = \frac{\int {\rm d}\vec{v}_1 {\rm d}\vec{v}_2 f(\vec{v}_1) f(\vec{v}_2) |\vec{v}_1 - \vec{v}_2|}{\left( \int {\rm d}\vec{v} f(\vec{v}) \right)^2}$, along with $n=\int{\rm d}\vec{v}f(\vec{v})$,
\begin{equation}
    R = \frac{1}{2} n^2 \Sigma v .
\end{equation}
$R$ is the total rate per unit volume, so the rate for a single particle is $\nsv$.
}

\section{Comparing mass precession and granularity in SMBH environments}
\label{appendix:prec_vs_relax}

An orbital shuffling effect we have not taken into account in examining SMBH environments (subsection~\ref{subsec:SMBH}) is granularity -- individual tugs from stars.
As discussed in subsection~\ref{subsec:random_walk_refresh}, this is a random process so the refresh time it induces is
\begin{equation}
\label{eq:gran_diff_limit}
    t_\text{ref}(\text{gran}) = \parfrac{D}{R}^2 t_\text{relax} ,
\end{equation}
as long as the effect of a single gravitational tug is less than $D/R$.

\REVDEL{From here on, this appendix is completely rewritten.}

\REVNEW{
In the cases we are interested in, the black hole mass is much greater than the mass of the stars around it $\MBH \gg M(r)$, otherwise precession would make the regular $\nsv$ rate correct.
In such cases, it was shown in \cite{rauch1996} that orbit parameters (specifically, angular momentum) change due to resonant relaxation according to
\begin{equation}
    \delta \sim \begin{cases}
        \frac{\sqrt{\bar{m} M(r)}}{\MBH} \frac{t}{T}
        & t \ll t_\text{prec} \\
        \frac{\sqrt{\bar{m} M(r)}}{\MBH} \parfrac{t_\text{prec} t}{T^2}^{1/2}
        & t \gg t_\text{prec}
    \end{cases}.
\end{equation}
}

\REVNEW{
We consider resonant relaxation since it is stronger than non-resonant relaxation;
therefore, when we show that mass precession is stronger than resonant relaxation, the result is also valid for non-resonant relaxation.
}

\REVNEW{
Using the expression for mass precession $t_\text{prec} = \frac{\MBH}{M(r)} T$,
\begin{equation}
    \delta \sim \begin{cases}
        \frac{\sqrt{\bar{m} M(r)}}{\MBH} \frac{t}{T}
        & \frac{t}{T} \ll \frac{\MBH}{M(r)} \\
        \sqrt{\frac{\bar{m}}{\MBH}} \parfrac{t}{T}^{1/2}
        & \frac{t}{T} \ll \frac{\MBH}{M(r)}
    \end{cases}.
\end{equation}
}

\REVNEW{
At the transition from linear growth to random walk, $\delta \sim \sqrt{\frac{\bar{m}}{M(r)}} = N^{-1/2}$.
The refresh time due to relaxation is when $\delta = \frac{D}{r}$;
comparing it with the refresh time due to mass precession $t_\text{ref}^\text{mass} = \frac{D}{r} \frac{\MBH}{N \bar{m}} T$:
\begin{equation}
    \frac{t_\text{ref}^\text{relax}}{t_\text{ref}^\text{mass}} = \begin{cases}
        N^{1/2}
        & \frac{D}{r} \ll N^{-1/2} \\
        \frac{D}{r} N
        & \frac{D}{r} \gg N^{-1/2}
    \end{cases}.
\end{equation}
Whatever the value of $D/r$ compared to $N^{-1/2}$, mass precession refresh time is shorter than the resonant relaxation refresh time.
Therefore, there is no need to take this effect into account if mass precession is already included.
}

\bibliography{references}{}
\bibliographystyle{aasjournal}



\end{document}